\documentclass[twocolumn, times]{aastex631}

\usepackage{xspace}
\usepackage{graphicx}
\usepackage{booktabs}
\usepackage[enable]{easy-todo}
\usepackage{xcolor}

\newcommand{\lya}{Lyman-$\alpha$\xspace}  
\newcommand{\tcm}{21$\,$cm\xspace}  

\newcommand{\centfig}[2][1]{
    \centering
    \makebox[\linewidth][c]{\includegraphics[width=#1\linewidth]{#2}}
}

\newcommand{\sjffixed}[1]{{\color{purple}\textbf{*}}}

\newcommand{\sethfixed}[1]{{\color{blue}\textbf{*}}}

\usepackage{amsmath}

\newcommand{\dlya}{\ensuremath{\delta_\alpha}}

\renewcommand{\vec}[1]{{\boldsymbol{\mathbf{#1}}}}

\newcommand{\uvec}[1]{\hat{\vec{#1}}}

\renewcommand{\la}{\ensuremath{\left\langle}}  
\newcommand{\ra}{\ensuremath{\right\rangle}}

\newcommand{\lv}{\ensuremath{\left\lvert}}   
\newcommand{\rv}{\ensuremath{\right\rvert}}

\newcommand{\ls}{\ensuremath{\left[}}        
\newcommand{\rs}{\ensuremath{\right]}}

\newcommand{\lp}{\ensuremath{\left(}}        
\newcommand{\rp}{\ensuremath{\right)}}

\newcommand{\degree}{$^{\circ}$}

\begin{document}

\title{A Detection of Cosmological \tcm Emission from CHIME in Cross-correlation
with eBOSS Measurements of the \lya Forest}

\date{\today}

\newcommand{\UBC}{Department of Physics and Astronomy, University of British Columbia, Vancouver, BC, Canada}
\newcommand{\MITP} {Department of Physics, Massachusetts Institute of Technology, Cambridge, MA, USA}
\newcommand{\MITK} {MIT Kavli Institute for Astrophysics and Space Research, Massachusetts Institute of Technology, Cambridge, MA, USA}
\newcommand{\TRU}{Department of Physical Sciences, Thompson Rivers University, Kamloops, BC, Canada}
\newcommand{\PI}{Perimeter Institute for Theoretical Physics, Waterloo, ON, Canada}
\newcommand{\DRAO}{Dominion Radio Astrophysical Observatory, Herzberg Astronomy \& Astrophysics Research Centre, National Research Council Canada, Penticton, BC, Canada}
\newcommand{\UBCO}{Department of Computer Science, Math, Physics, and Statistics, University of British Columbia-Okanagan, Kelowna, BC, Canada}
\newcommand{\McGill}{Department of Physics, McGill University, Montreal, QC, Canada}
\newcommand{\UofTastro}{David A.\ Dunlap Department of Astronomy \& Astrophysics, University of Toronto, Toronto, ON, Canada}
\newcommand{\UofTphys}{Department of Physics, University of Toronto, Toronto, ON, Canada}
\newcommand{\WVU} {Department of Computer Science and Electrical Engineering, West Virginia University, Morgantown WV, USA}
\newcommand{\WVUA} {Department of Physics and Astronomy, West Virginia University, Morgantown, WV, USA}
\newcommand{\WVUGWAC} {Center for Gravitational Waves and Cosmology, West Virginia University, Morgantown, WV, USA}
\newcommand{\Yale}{Department of Physics, Yale University, New Haven, CT, USA}
\newcommand{\YaleA}{Department of Astronomy, Yale University, New Haven, CT, USA}
\newcommand{\Dunlap}{Dunlap Institute for Astronomy and Astrophysics, University of Toronto, Toronto, ON, Canada}
\newcommand{\RRI}{Raman Research Institute, Sadashivanagar,   Bengaluru, India}
\newcommand{\ASIAA}{Institute of Astronomy and Astrophysics, Academia Sinica, Taipei, Taiwan}
\newcommand{\CITA}{Canadian Institute for Theoretical Astrophysics, Toronto, ON, Canada}
\newcommand{\CIFAR}{Canadian Institute for Advanced Research,  Toronto, ON, Canada}
\newcommand{\WVUphysastro}{Department of Physics and Astronomy, West Virginia University, Morgantown, WV, USA}
\newcommand{\ASU}{Department of Physics, Arizona State University, Tempe, AZ, USA}
\newcommand{\TSI}{Trottier Space Institute, McGill University, Montreal, QC, Canada}
\newcommand{\KAI}{Kapteyn Astronomical Institute, University of Groningen, Groningen, The Netherlands}

\shortauthors{CHIME Collaboration}
\collaboration{100}{The CHIME Collaboration:}

\author[0000-0001-6523-9029]{Mandana Amiri}
\affiliation{\UBC}
\author[0000-0003-3772-2798]{Kevin Bandura}
\affiliation{\WVU}
\affiliation{\WVUGWAC}
\author[0000-0002-7758-9859]{Arnab Chakraborty}
\affiliation{\McGill}
\affiliation{\TSI}
\author[0000-0001-7166-6422]{Matt Dobbs}
\affiliation{\McGill}
\affiliation{\TSI}
\author[0000-0002-6899-1176]{Mateus Fandino}
\affiliation{\TRU}
\author[0000-0002-0190-2271]{Simon Foreman}
\affiliation{\ASU}
\author[0000-0002-4903-838X]{Hyoyin Gan}
\affiliation{\UBC}
\author[0000-0002-1760-0868]{Mark Halpern}
\affiliation{\UBC}
\author[0000-0001-7301-5666]{Alex S. Hill}
\affiliation{\UBCO}
\affiliation{\DRAO}
\author[0000-0002-4241-8320]{Gary Hinshaw}
\affiliation{\UBC}
\author[0000-0003-4887-8114]{Carolin H\"ofer}
\affiliation{\UBC}
\affiliation{\KAI}
\author[0000-0003-1455-2546]{T.L. Landecker}
\affiliation{\DRAO}
\author[0000-0002-0309-9750]{Zack Li}
\affiliation{\CITA}
\author[0000-0001-8064-6116]{Joshua MacEachern}
\affiliation{\UBC}
\author[0000-0002-4279-6946]{Kiyoshi Masui}
\affiliation{\MITK}
\affiliation{\MITP}
\author[0000-0002-0772-9326]{Juan Mena-Parra}
\affiliation{\Dunlap}
\affiliation{\UofTastro}
\author[0000-0001-8292-0051]{Nikola Milutinovic}
\affiliation{\UBC}
\author[0000-0002-2626-5985]{Arash Mirhosseini}
\affiliation{\UBC}
\author[0000-0002-7333-5552]{Laura Newburgh}
\affiliation{\Yale University}
\author[0000-0002-2465-8937]{Anna Ordog}
\affiliation{\UBCO}
\affiliation{\DRAO}
\author[0000-0002-8671-2177]{Sourabh Paul}
\affiliation{\McGill}
\affiliation{\TSI}
\author[0000-0003-2155-9578]{Ue-Li Pen}
\affiliation{\CITA}
\author[0000-0002-9516-3245]{Tristan Pinsonneault-Marotte}
\affiliation{\UBC}
\author[0000-0001-6967-7253]{Alex Reda}
\affiliation{\Yale University}
\author[0000-0002-4543-4588]{J. Richard Shaw}
\affiliation{\UBC}
\author[0000-0003-2631-6217]{Seth R. Siegel}
\affiliation{\McGill}
\affiliation{\TSI}
\affiliation{\PI}
\author[0000-0003-4535-9378]{Keith Vanderlinde}
\affiliation{\UofTastro}
\affiliation{\Dunlap}
\author[0000-0002-1491-3738]{Haochen Wang}
\affiliation{\MITK}
\affiliation{\MITP}
\author[0000-0002-6669-3159]{D. V. Wiebe}
\affiliation{\UBC}
\author[0000-0001-7314-9496]{Dallas Wulf}
\affiliation{\McGill}
\affiliation{\TSI}

\correspondingauthor{T.~Pinsonneault-Marotte}
\email{tristpinsm@phas.ubc.ca}

\graphicspath{{./}{figures/}}

\begin{abstract}
    We report the detection of \tcm emission at an average redshift $\bar{z} =
    2.3$ in the cross-correlation of data from the Canadian Hydrogen Intensity
    Mapping Experiment (CHIME) with measurements of the \lya forest from eBOSS.
    Data collected by CHIME over 88 days in the $400-500$~MHz frequency band ($1.8
    < z < 2.5$) are formed into maps of the sky and high-pass delay filtered to
    suppress the foreground power, corresponding to removing cosmological scales
    with $k_\parallel \lesssim 0.13\ \text{Mpc}^{-1}$ at the average redshift.
    Line-of-sight spectra to the eBOSS background quasar locations are extracted
    from the CHIME maps and combined with the \lya forest flux transmission
    spectra to estimate the \tcm-\lya cross-correlation function. Fitting a
    simulation-derived template function to this measurement results in a
    $9\sigma$ detection significance. The coherent accumulation of the signal
    through cross-correlation is sufficient to enable a detection despite excess
    variance from foreground residuals $\sim6-10$ times brighter than the
    expected thermal noise level in the correlation function. These results are
    the highest-redshift measurement of \tcm emission to date, and set the stage
    for future \tcm intensity mapping analyses at $z>1.8$.
\end{abstract}

\section{Introduction}
\label{sec:intro}

Emission from the hyperfine transition of neutral hydrogen (HI) at \tcm can be
used to efficiently map the large-scale structure (LSS) of the Universe over
most of its history, a technique known as Hydrogen intensity mapping. In this
approach, HI contained in galaxies or the inter-galactic medium is detected in
aggregate -- integrated over the relatively coarse angular resolution afforded
by telescopes at radio wavelengths, where redshifted \tcm emission is measured.
Such telescopes can be designed to cost-effectively observe large sky areas at
high sensitivity, and digital receivers enable broad bandwidths to be sampled
with fine frequency resolution, which directly maps to redshift due to the
monochromaticity of the signal \citep{2009petersonHIM}.

Over the last decade, instruments have been built with the aim of measuring this
signal in the late Universe (e.g. CHIME, MeerKAT) as well as at higher redshift
(HERA, EDGES, SARAS, PAPER, among others). Published results from the field
include upper limits \citep{2022ApJ_hera}, detections in cross-correlation
\citep{pen2009,chang2010,masui2013,anderson2018,tramonte2020,li2021-parkesxwigglez,2022MNRAS_wolz_gbt,chime_collaboration_detection_2022,2023MNRAS_meerKAT_xcorr},
and a first detection in auto-correlation \citep{2023arXiv_meerKAT}. The main
challenge for all of these efforts is separating the faint \tcm signal from the
extremely bright foreground emission of our galaxy and radio point sources,
which can in principle be achieved by making use of their different spectral
properties, but is complicated by instrumental effects and radio-frequency
interference (RFI). Until foreground separation methods mature to a point where
the signal becomes dominant, cross-correlation with external surveys has proven
to be an effective way of mitigating residual foregrounds and obtaining
interesting scientific constraints, as demonstrated by the recently reported
detection from CHIME in cross-correlation with SDSS galaxy surveys
\citep{chime_collaboration_detection_2022}.

The Canadian Hydrogen Intensity Mapping Experiment (CHIME)
\citep{the_chime_collaboration_overview_2022} is a compact interferometer
composed of four cylindrical reflectors instrumented with a total of 1024
dual-polarisation antennas. It is located at the Dominion Radio Astrophysical
Observatory (DRAO) in Penticton, Canada, where it observes the entire sky
visible at latitude $\sim 49$\degree, operating as a driftscan telescope over
1024 channels in the $400-800$~MHz band (corresponding to \tcm emission
redshifted by $2.5 > z > 0.8$). CHIME has been operating continuously since
2017.

The redshift range available to CHIME overlaps with measurements of the \lya
forest from eBOSS \citep{2020ApJ...901..153D}, allowing for a cross-correlation
analysis to be carried out in the high-redshift end of the CHIME band. At these
redshifts, the quasar catalogue that was stacked on in
\citet{chime_collaboration_detection_2022} is sparser, and there is more
statistical weight in the \lya data. \lya forest measurements are characterized
by absorption of quasar light by HI clouds along the line of sight whereas the
\tcm signal is in emission, so we expect the cross-correlation with CHIME maps
to be {\em negative} at small separations, which provides a unique signal that
is difficult to mimic any other way. This is a feature that has been highlighted
by \cite{2017_carucci_xcorr}, who also emphasize the usefulness of such a
cross-correlation to mitigate foreground contamination and break degeneracies in
a power spectrum analysis.

Several studies have found evidence that various classes of galaxies are
correlated with \lya absorption by the IGM up to several tens of megaparsecs
\citep{2017ApJ...835..281M,2021ApJ...909..117M,2021ApJ...907....3L}, but these
studies have been limited to small numbers of galaxies and/or absorption
systems. A joint analysis of CHIME and eBOSS \lya forest data will provide a
huge statistical improvement on both sides, opening a new window on the
connection between HI-rich galaxies and low-density HI in the surrounding IGM,
and thus shedding light on the role of HI in galaxy evolution (e.g.\ the
strength of HI inflows from the IGM onto denser systems) when the cosmic star
formation rate is at its peak.

In this work, we present a detection of \tcm emission in the cross-correlation
of CHIME data and eBOSS \lya forest measurements, but leave the interpretation
of the signal (including modelling of its cosmological and astrophysical
implications) for future work. We describe the data used in this analysis in
Section~\ref{sec:data} and the CHIME processing pipeline in
Section~\ref{sec:proc}. Section~\ref{sec:analysis} explains the
cross-correlation method we use. A model for the signal and its inclusion in
simulations are described in Section~\ref{sec:sims}. The results of the analysis
are presented in Section~\ref{sec:results}, and some validation tests in
Section~\ref{sec:valid}. Finally, we conclude in Section~\ref{sec:conc}.

\section{Data}
\label{sec:data}

\subsection{eBOSS}

The \lya forest measurements we use in our analysis are those from the eBOSS
sixteenth data release (SDSS DR16) \citep{2020ApJ...901..153D}. The data products
consist of fractional flux transmission along the line of sight to backlight
quasars, defined as
\begin{equation}
    \delta_q(\lambda) = \frac{f_q(\lambda)}{\bar{F}(\lambda)C_q(\lambda)} - 1\text{,}
\end{equation}
where $f_q$ is the measured flux, $C_q$ is the unabsorbed quasar continuum and
$\bar{F}$ is the mean transmission. The product $\bar{F}(\lambda)C_q(\lambda)$
was modelled and fit to the data as explained in \citet{2020ApJ...901..153D}.
Apart from interpolating onto the CHIME frequency band
(Section~\ref{sec:frequency_remapping}), no additional processing is performed
on the catalogue provided by eBOSS.

\begin{figure}
    \centfig{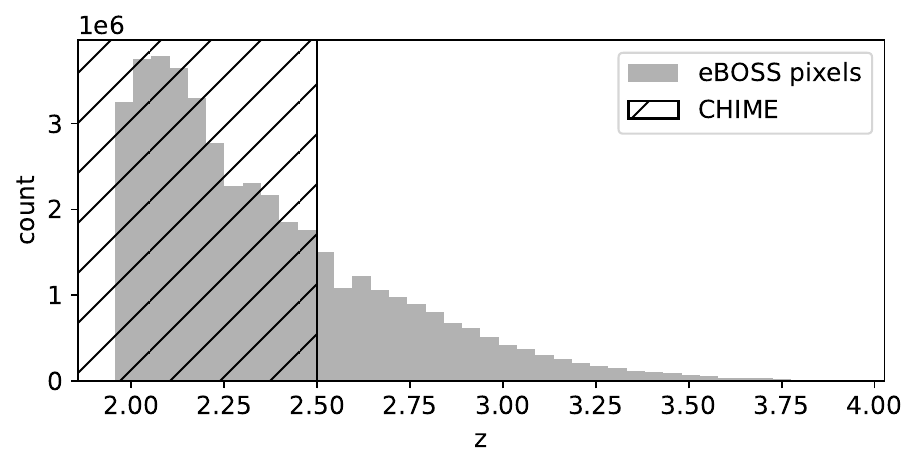}
    \caption{Histogram of redshifts of eBOSS \lya forest samples and their
    overlap with the redshift-range probed by CHIME.}
    \label{fig:eboss_pix}
\end{figure}

There are 210005 background quasars in the eBOSS \lya catalogue, with redshifts
$2.1 < z < 4$. The CHIME data go down to 400 MHz, corresponding to \tcm
radiation redshifted at $z=2.5$, which overlaps with a total of
$\sim3\times10^7$ spectral samples in the eBOSS measurements (see
Figure~\ref{fig:eboss_pix}). We include the full eBOSS spectra in this total and
the analysis, including the regions blueward of the Lyman-$\beta$ wavelength
($\sim102$~nm)
at the quasar redshift (referred to as the Ly$\beta$ region in the eBOSS
papers). This part of the spectrum is treated separately in their analysis
because it can include absorption from both transition levels, but since we are
cross-correlating with an external dataset, there is no concern of contamination
from Ly$\beta$ correlations. Damped \lya systems are identified and masked in
the eBOSS data products, so they do not contribute to the cross-correlation with
CHIME. The quasars are distributed throughout the SDSS North and South Galactic
Cap fields (NGC and SGC), which are fully contained within the sky area observed
by CHIME.

\subsection{CHIME}

The CHIME data that was used in this work is composed of nighttime observations
from 88 days spanning the calendar year 2019. This amounts to $\sim 1000$~hours
of total observation time, distributed approximately uniformly over the sidereal
day. The full frequency range observed with CHIME is $400-800$~MHz but we use only
the lowest quarter of the band, i.e. $400-500$~MHz or $z > 1.8$, that has overlap
with the \lya forest measurements. This is a similar set of observations to what
was used in \citet{chime_collaboration_detection_2022} to detect \tcm emission
in cross-correlation with SDSS galaxy catalogues (refer to that paper and
citations therein for a general description of the CHIME data). However, that
analysis considered data from the upper part of the frequency band,
$585-800$~MHz, whereas this work is the first report on the high-redshift end.

\section{Processing}
\label{sec:proc}

The processing that the CHIME data undergoes prior to cross-correlation with the
\lya forest is largely the same as described in the first detection paper
(Section~3 of \citealt{chime_collaboration_detection_2022}). In this section we
summarise the main steps and highlight improvements that were made to the
pipeline since the publication of that analysis as well as aspects of the
processing that had to be modified to accommodate the lower frequency band.

\subsection{Sidereal Stacking Pipeline}

The final product of the CHIME processing pipeline is a set of visibilities for
unique baselines measuring the sky on a grid of right ascension spanning the
sidereal day. To produce this ``sidereal stack'', observations from individual
days are collected and passed through daily processing that includes RFI
flagging, gain calibration and regridding from local time to a fixed grid in
sidereal time. Each day is inspected visually via a set of data quality metrics
to identify and flag days that appear corrupted. The remaining days are averaged
together to form the final sidereal stack that is used in this analysis. Further
details on each stage of this processing can be found in
\cite{chime_collaboration_detection_2022}.

\subsubsection{Improved Flagging}

Since the publication of the first CHIME detection, some improvements to the
processing pipeline have been made. These include:

\paragraph{Narrowband gain errors} The calibration algorithms employed in the
real-time pipeline attempt to identify and mask RFI-like features in the
underlying data prior to estimating the frequency- and feed-dependent gains.
However, these algorithms are not entirely robust to RFI and we have found that
transient RFI occasionally biases the resulting gain estimates.  This is true
for both the ``digital gains'', which are updated semi-annually and whose
purpose is to minimize quantization noise when truncating the data streams to 4
bits (real) + 4 bits (imaginary), and the ``calibration gains'', which are
updated daily and whose purpose is to correct instability in the analog receiver
chain.  The offending RFI is narrowband, usually with a bandwidth less than a
single 390.625~kHz frequency channel, and results in spikes in the estimated
gain as a function of frequency. These spikes are imprinted on the foregrounds
when the gains are applied to the visibility data, leaking foreground power to
high delays (the conjugate axis to frequency, where smooth foregrounds would
otherwise be confined to low delays).

We have developed a new method to search the archived gains for these narrowband
artifacts and then mask the corresponding frequencies and times in the
visibility data.  For each time sample in the visibility data, we calculate the
product of the digital gain and calibration gain that was applied to each feed
and frequency by the real-time pipeline, average the amplitude over feeds, and
then perform a search for narrowband features along the frequency axis.  The
search algorithm applies an aggressive, high-pass filter along the frequency
axis, masks the frequency channel that is the largest outlier, reapplies the
filter accounting for the newly masked data, and iterates until all unmasked
frequencies are less than $6 \sigma$, where $\sigma$ is an estimate of the
standard deviation of the noise.  This procedure masked 6\% of the data that was
considered for this analysis.

\paragraph{Decorrelation events} On rare occasions, communication errors during
the corner-turn operation can cause certain data streams in the correlator to
become misaligned with all other data streams.  This results in one quarter of
the feeds decorrelating with all other feeds for 64 of the 1024 frequency
channels, which persists until the data streams are re-aligned by restarting the
correlator.  The 64 decorrelated frequency channels are uniformly spaced across
the band and leak significant foreground power to high delays.  We now perform
an automated search for these decorrelation events and exclude any frequency
channel and time that is affected from further analysis.  In total, 0.07\% of
the data considered for this analysis was masked for this reason, corresponding
to a single decorrelation event that persisted for the majority of one sidereal
day.

\paragraph{Excessively small weights} The real-time RFI flagging excises corrupt
samples within the time integrations that are eventually saved to disk; for
integrations where this excision is nearly total, the data saved to disk can be
extremely noisy. A similar effect can occur due to packet loss or other digital
errors in the X-engine. These integrations are assigned very low noise weights
in subsequent analysis, but if the issues are confined to very narrow frequency
bands or small subsets of baselines, the low weights were not properly
incorporated in previous versions of the processing. This has been corrected
using an extra flagging step in the offline data pipeline, which flags 0.5\% of
all time-frequency samples used in this analysis.

\subsubsection{Thermal Gain Correction}

Section 3.2.4 of \cite{chime_collaboration_detection_2022} describes a stage of
the data processing that uses measurements of the outside temperature to correct
common-mode thermal variations in the amplitude of the gains of the analog
receiver chains.  We recently discovered two errors in the resulting instrument
stability as reported in \cite{chime_collaboration_detection_2022}.  First, the
pre(post)-correction stability of $0.8\%$ ($0.5\%$) was reported as the standard
deviation in fractional power, but it is actually the standard deviation in
fractional voltage.  Second, the software that was deployed in the pipeline
mistakenly applied the inverse of the thermal correction, which amplified the
thermal variations instead of mitigating them.  We have fixed the software error
and applied the correct factor to the data used for this analysis.  As a result,
the estimated stability for the analysis presented in
\cite{chime_collaboration_detection_2022} is $2.7\%$ (standard deviation in
fractional power), while the estimated stability for this analysis is $1.0\%$.

\subsubsection{Impact on Previously-reported Results}

The changes to the CHIME processing pipeline described in this section do not
result in an improved signal-to-noise on the detection of cosmological \tcm
emission in cross-correlation with the eBOSS galaxy catalogues that was reported
in \citet{chime_collaboration_detection_2022}. In fact, the noise level in the
more recent revision of the processed data is found to be elevated compared to
what was measured in the previous one, but this has no qualitative impact on the
results reported in \citet{chime_collaboration_detection_2022}. Investigating
the impact of these changes and continued iteration on the processing pipeline
is currently ongoing.

\subsection{Ringmaps}

\begin{figure}
    \centfig[1.1]{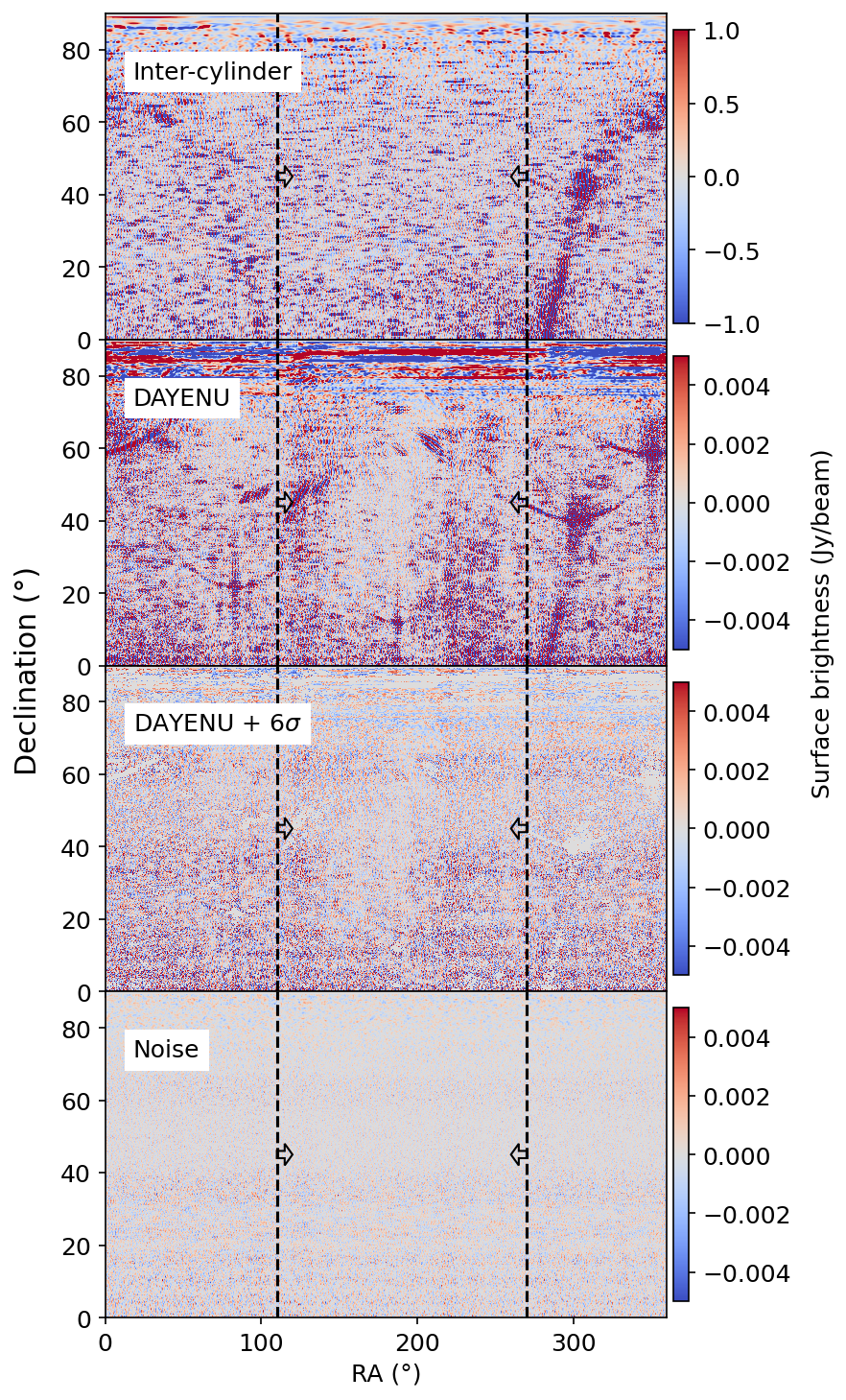}
    \caption{ \emph{(First)} Ringmap generated from inter-cylinder baselines
        ($\gtrsim 22$~m East-West distance) at 476.56~MHz for the XX
        polarisation. \emph{(Second)} Ringmap generated from visibilities
        filtered using DAYENU with a 200~ns delay cut. \emph{(Third)} The same
        as the second panel but with an outlier mask derived from a $6\sigma$
        threshold based on the expected noise. \emph{(Fourth)} A noise
        realisation at the level expected in the CHIME data, delay-filtered in
        the same way as in the previous panels. In all panels, vertical dashed
        lines indicate the region that was used to estimate the delay power
        spectrum in Figure~\ref{fig:delay_spec}. }
    \label{fig:rmap}
\end{figure}

The way CHIME observes the sky -- a primary beam that is a few degrees wide in
the East-West direction but spans the South and North horizons, illuminating a
regularly-spaced grid of feeds as the sky drifts overhead -- is amenable to a
particular map-making method, which produces ``ringmaps''. At every time sample,
corresponding to a specific right ascension (RA) on meridian, the North-South
(NS) baselines are phased and combined so as to form a grid of synthesized beams
within the instantaneous field of view. As the sky drifts overhead, each formed
beam records its intensity along a ring of constant declination. The image
generated in this way is called a ringmap. An example is displayed in the top
panel of Figure~\ref{fig:rmap}. A detailed description of this process is given
in \citet{chime_collaboration_detection_2022}.

For the cross-correlation analysis, we generate ringmaps from the sidereal
stacks at each polarisation, and find the angular pixel that contains the
position of each quasar in the eBOSS catalog. The spectrum from this pixel is
what will be correlated with the corresponding \lya forest spectrum.

\subsection{Lower Band Processing}

\subsubsection{Beam Calibration}

\begin{figure*}
    \centfig{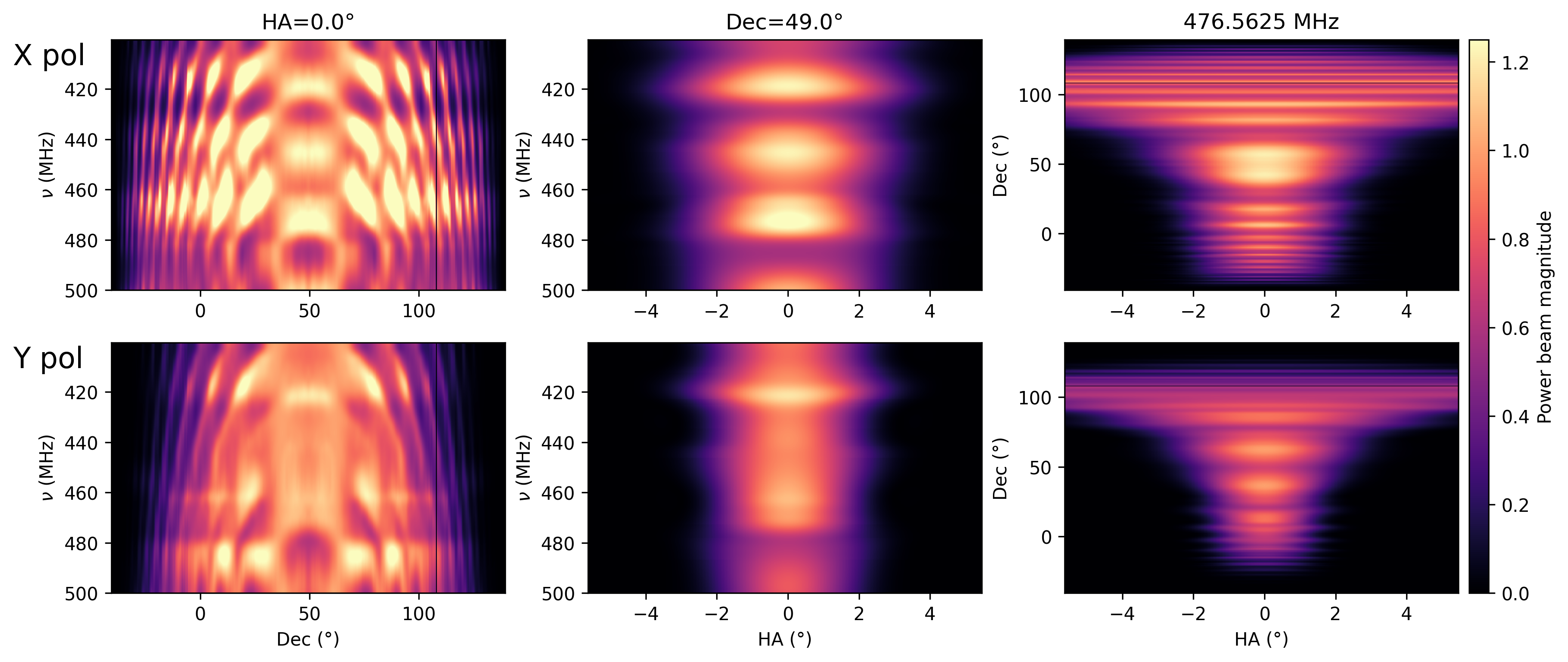}
    \caption{Slices of the 3D beam model, from left to right, at constant hour
    angle, declination, and frequency. The top row is the power beam for the X
    polarisation, and the bottom row is Y. The beam pattern has been projected
    onto sky coordinates HA and declination, which is why it broadens
    significantly near the NCP. Note that the declination axis has been extended
    beyond 90\degree to reflect the sensitivity of the beam to regions of the
    sky on either side of the NCP.}
    \label{fig:beam}
\end{figure*}

When ringmaps are generated, a beam model is used to deconvolve the effect of
the main lobe of the primary beam. The model that was used in
\citet{chime_collaboration_detection_2022} was derived from data by fitting to
the expected flux of a large number of radio point sources. This model has since
been extended to the lower band considered in this analysis, using the same
methods that were described in detail in
\citet{chime_collaboration_detection_2022}. Slices along each axis of the beam
model are presented in Figure~\ref{fig:beam}, for both polarisations.

\subsubsection{Frequency Flagging}

Although RFI is flagged algorithmically at multiple stages in the real-time and
daily processing, low-level events which are fainter than the foregrounds are
very hard to detect. This contamination becomes apparent after a high-pass delay
filter has been applied to remove most of the foreground power, and is strong
enough to obscure the cosmological signal. We identify affected frequencies by
iteratively filtering and flagging outlier frequencies until none remain. The
flagging was performed by visually inspecting maps generated from the filtered
data at every frequency, where the spatial information helps identify bad cases.
In the end, 47\% of the data in the $400-500$~MHz band was flagged in this way, in
addition to persistent RFI bands that are statically masked, such that a total
of 65\% of the band was masked in this analysis.

We attributed this contamination to RFI, but instrumental miscalibration such as the
narrowband gain errors described above could have a similar effect. As the
calibration and flagging parts of the daily pipeline continue to improve, we may
be able to recover some of these flagged frequencies.

\subsubsection{Delay Filter}

In this analysis we suppress the foregrounds by high-pass filtering the
$400-500$~MHz band in delay. The filter is implemented using the DAYENU method
to account for the masked regions of the frequency band \citep{dayenu}. We set
the lower edge of the delay pass-band at 200~ns, which was chosen to encompass
the bulk of foreground power at all declinations in the delay power spectrum, as
illustrated in Figure~\ref{fig:delay_spec}. This power spectrum was evaluated
using the Gibbs sampling method described in Appendix~A of
\cite{chime_collaboration_detection_2022} and estimating the variance across
right ascensions 110-270\degree, a region corresponding roughly to the eBOSS NGC
field, where the majority of the \lya quasars are found. Shifting this threshold
by 50 or 100~ns in either direction did not appear to improve the
signal-to-noise, but no further optimisation was attemped. Filtering delays
below 200~ns corresponds to removing large scales with $k_\parallel \lesssim
0.13\ \text{Mpc}^{-1}$, at the average redshift $\bar{z} = 2.3$ ($\sim
430$~MHz). Figure~\ref{fig:rmap} shows maps at a single frequency and
polarisation before and after the delay filter. Pixels near bright foreground
features tend to remain corrupted after filtering, so we also derive a mask that
excises any pixel above a $6\sigma$ threshold determined by the expected thermal
noise. This threshold is what was used in
\citet{chime_collaboration_detection_2022} and is sufficient but necessary to
enable a detection of the \tcm signal.

\begin{figure}
    \centfig{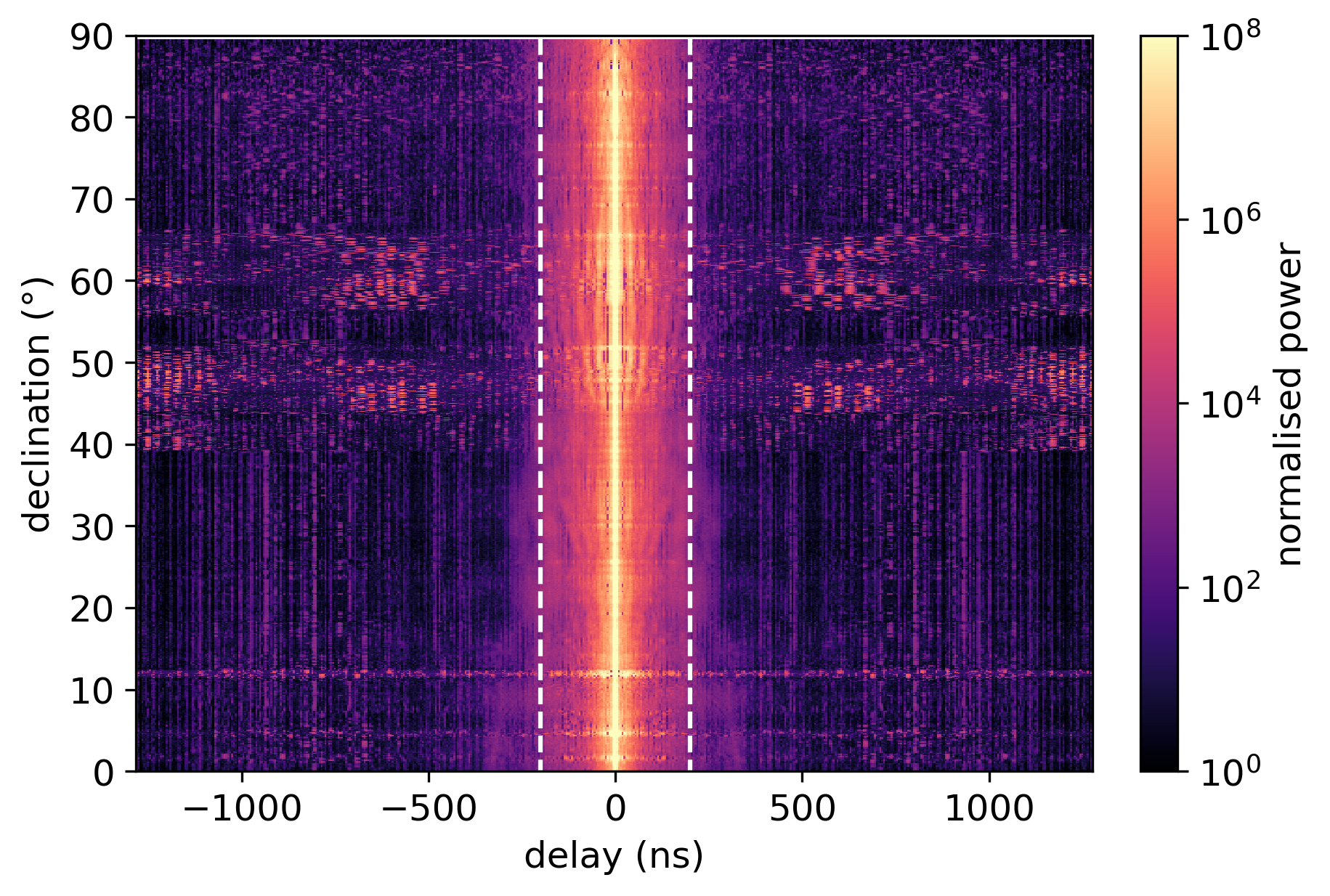}
    \caption{Delay power spectrum of a CHIME ringmap evaluated across right ascensions between 110-270\degree, at fixed declination. The colour scale represents power in units of noise power. Vertical dashed lines show the lower bounds of the DAYENU filter pass-band. Significant contamination exists at all delays. Some of the horizontal features can be attributed to bright point sources in the maps, but the vertical lines are of unknown origin.}
    \label{fig:delay_spec}
\end{figure}

\section{Cross-correlation Method}
\label{sec:analysis}

The cross-correlation function of the \tcm temperature $T_{21}$ and the \lya
forest relative flux transmission $\dlya$ is defined at a given redshift as
\begin{equation}
    \xi(\vec{r}, \vec{r}') = \la T_{21}\lp\vec{r}\rp \dlya\lp\vec{r'}\rp \ra
    \text{,}
\end{equation}
where $\vec{r}$ and $\vec{r}'$ are three-dimensional comoving positions and the
angle brackets indicate an ensemble average. Assuming statistical isotropy, the
correlation function only depends on the magnitude of the separation in comoving
space: $\xi(\vec{r}, \vec{r}') = \xi(\lv\vec{r} - \vec{r}'\rv)$. However,
measurements of these fields are located on our past lightcone -- in redshift
space and on the sphere -- denoted by $(z, \uvec{n})$, so this is where we will
evaluate the correlation function.

In aggregate, the line-of-sight measurements from eBOSS and CHIME contain
three-dimensional information on the two fields, i.e. one could evaluate the
correlations along the redshift ($z$) and angular ($\uvec{n}$) coordinates.
However, CHIME's resolving power is much greater in frequency than it is on the
sky, so we expect the signal-to-noise to be stronger along this axis. Given the
significant added complexity of a 3-d analysis, in this work we only consider
correlations along the line-of-sight.

Here, we will describe the procedure used to compute an estimate of the radial
cross-correlation function in redshift space, corresponding to
\begin{equation}
    \xi\lp\Delta z\rp = \la
        \int dz~T_{21}^i\lp z\rp \dlya^i\lp z + \Delta z\rp
    \ra_i
    \text{,}
\end{equation}
where the angle brackets denote the average over individual line-of-sight
observations labelled by $i$. We describe how redshifts are computed for each
observable in Sec.~\ref{sec:frequency_remapping}, and how the cross-correlation
is computed in Sec.~\ref{sec:correlation_function_estimation}.

\subsection{Frequency Remapping}
\label{sec:frequency_remapping}

The eBOSS \lya forest spectra measurements are arranged in wavelength bins
uniformly spaced in $\ln\lambda$ and spanning $\sim 360-700$~nm, whereas the
CHIME data is uniformly spaced in frequency in the $400-800$~MHz band. In order
to correlate the two, it will be necessary to map these onto a common grid in
redshift. Given that the resolution of the CHIME data is coarser by a factor of
$\sim 1.2-1.5$ in the redshift range where they overlap, and that it will be
convenient to work with the CHIME data directly, we choose to map the \lya data
onto the CHIME band. The mapping is defined by matching the frequency of \lya
absorption to the observed frequency of \tcm emission for the corresponding
redshift, a relationship given by the ratio of their rest wavelengths
\begin{eqnarray}
    \nu' = \frac{\lambda_\alpha}{\lambda_{21}} \nu \text{.}
\end{eqnarray}

We use the reverse Lanczos interpolation method described in Section~3 of
\citet{chime_collaboration_detection_2022} to regrid the \lya forest spectra
onto the CHIME band. The regions of the spectra that are empty (redshifts larger
than that of the background quasar, or otherwise masked) are set to zero in the
regridded data.

\subsection{Correlation Function Estimation}
\label{sec:correlation_function_estimation}

We will write the discrete samples in frequency as $T_{21}[n] = T_{21}(n
\Delta\nu + 400 \mathrm{MHz})$, with integers $n$ labelling samples at the CHIME
frequency resolution, $\Delta\nu = 390.625$~kHz. Both the \lya and CHIME spectra
have noise estimates associated with every sample so we use these to weight the
samples and improve signal-to-noise. With inverse variance weights, $w =
\sigma^{-2}$, we estimate the cross-correlation function as
\begin{equation}
    \xi[n] = W^{-1} \sum_i^M \sum_j^N w_{21}^i[j] \, w_\alpha^i[j+n]
    \, T_{21}^i[j] \, \dlya^i[j + n] \: \text{,}
\end{equation}
where $i$ labels lines of sight, $n$ indexes separation in frequency, and
$W^{-1}$ is a normalisation term given by
\begin{equation}
    W = \sum_i^M \sum_j^N w_{21}^i[j] \, w_\alpha^i[j+n] \text{.}
\end{equation}
$M=210005$ is the total number of lines of sight and $N\leq 255$ is the number
of samples along the frequency axis being averaged together at that separation.

\section{Signal Model and Simulations}
\label{sec:sims}

To allow us to interpret the results of this analysis and assess detection
significance, we generate simulations of correlated \tcm and \lya forest
measurements. These are run through the same processing pipeline and analysis as
the data in order to forward model the cross-correlation result. Although the
models described below are parametric, we do not attempt to fit them to the
data. Instead, the purpose of these models is to generate a reasonably realistic
template to compare to our measurement and assess the S/N.

\subsection{Large-scale Structure Simulation}

A detailed description of the model used to generate maps of large-scale
structure in the CHIME redshift range is given in Section~5 of
\citet{chime_collaboration_detection_2022}. Very briefly, a realisation of
Gaussian fluctuations on the lightcone is generated at the desired redshifts
given a non-linear matter power spectrum evaluated using the halo model
prediction from \citet{mead2021}.

\subsection{CHIME Signal Model}

The HI linear bias, Fingers of God effect, and \tcm brightness temperature are
modelled to generate a map of \tcm temperature. See Section~5 of
\citet{chime_collaboration_detection_2022} for the model definitions. In this
work we fix the parameters to the fiducial values defined there. Synthetic CHIME
observations are derived from these maps, and directly substituted for the real
data in the cross-correlation analysis.

\subsection{\lya Signal Model}

In order to model the \lya forest measurements, we need a prescription for
converting the matter density fluctuation field $\delta_m(z, \uvec{n})$
generated by our LSS simulation into an optical depth to \lya photons. A
commonly used and straightforward approach is to model the mildly non-linear
baryon field as log-normally distributed \citep{bi_evolution_1997} and the \lya
absorption by the fluctuating Gunn-Peterson approximation (FGPA)
\citep{farr_lyacolore_2020}.

The log-normal transform is performed so as to preserve the variance of the
original field $\sigma_m^2$:
\begin{equation}
    1 + \delta_{LN}(z, \uvec{n}) = \exp\ls \delta_m(z, \uvec{n})
    - \frac{\sigma_m^2(z)}{2}\rs \text{.}
\end{equation}
This step also ensures that the density is strictly non-negative.

The optical depth $\tau$ is proportional to the number density of HI atoms,
which the FGPA assumes is related to the density by a power law
\begin{equation}
    \tau(z, \uvec{n}) = \tau_0(z) \lp 1 + \delta_{LN}(z, \uvec{n}) \rp^\alpha
    \text{.}
\end{equation}
We adopt the parameter values
\begin{equation}
    \tau_0(z) = 0.3 \lp \frac{1 + z}{3.4} \rp^{4.5}, \quad
    \alpha = 1.6 \text{,}
\end{equation}
quoted in \citet{cieplak_towards_2016} and \citet{seljak_bias_2012}.

These steps are illustrated in the top three panels of
Figure~\ref{fig:mock_steps} for a realisation of LSS along a given line of
sight.

Rather than use the FGPA, it might be more self-consistent to derive the optical
depth from the HI density that was modelled for the \tcm temperature directly.
This is expected to be part of a future effort to more carefully model the \lya
signal and perhaps derive constraints on the model parameters. We emphasize that
for the purpose of this work, the simplest approach was taken. However, we will
show that a simple FGPA-based model is sufficient to describe our measurements
at the current signal-to-noise level, so we leave more detailed modelling to
future analyses.

\subsection{Mock \lya Forest Catalogue}

\begin{figure}[htb!]
    \centfig{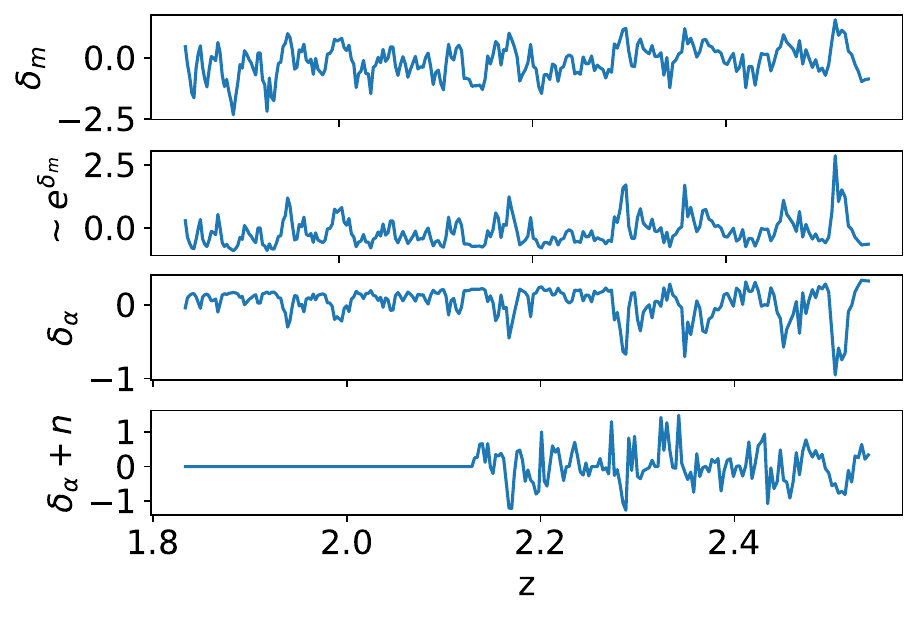}
    \caption{Fluctuations along a given line of sight in redshift, at subsequent
    steps of the procedure for generating simulated \lya forest spectra. From
    top to bottom: the matter fluctuations from the LSS simulation; the result
    of the log-normal transform; the Lyman-alpha fractional flux transmission;
    the final spectrum masked and with noise added to emulate the real
    data.}
    \label{fig:mock_steps}
\end{figure}

From the \lya fractional flux transmission simulation we extract individual
spectra along the lines of sight at a number of sky locations, chosen to be
those of the quasars in the eBOSS catalogue. Each spectrum is also masked so
that regions of missing data -- redshifts larger than the backlight quasar and
instrumental masks -- match those in the corresponding eBOSS data. Gaussian
random noise is optionally added to the simulated spectra at a level consistent
with the noise variance recorded alongside the data. This results in a mock
catalogue with exactly the same quasar sample as the data, and thus the same
sensitivity to cross-correlation with CHIME, but with a synthetic \lya forest
signal. The bottom panel of Figure~\ref{fig:mock_steps} shows a mock spectrum
generated in this way.

It must be noted that these simulations neglect a number of effects that would
be necessary to include in order to produce realistic simulations of the eBOSS
spectra (and which \emph{are} included in more sophisticated simulations, e.g.
\citet{farr_lyacolore_2020}). Among the most significant are
\begin{itemize}
    \item Redshift-space distorsions.
    \item Metal and Lyman-$\beta$ line absorption.
    \item High column density absorbers (which include damped \lya absorber systems)
    \item Biases due to the modelling of the quasar spectra continuum.
\end{itemize}
A careful accounting of these will be necessary for future work that will aim to
constrain the physics of the \lya forest.

\section{Results}
\label{sec:results}

\begin{figure*}[htb]
    \centfig{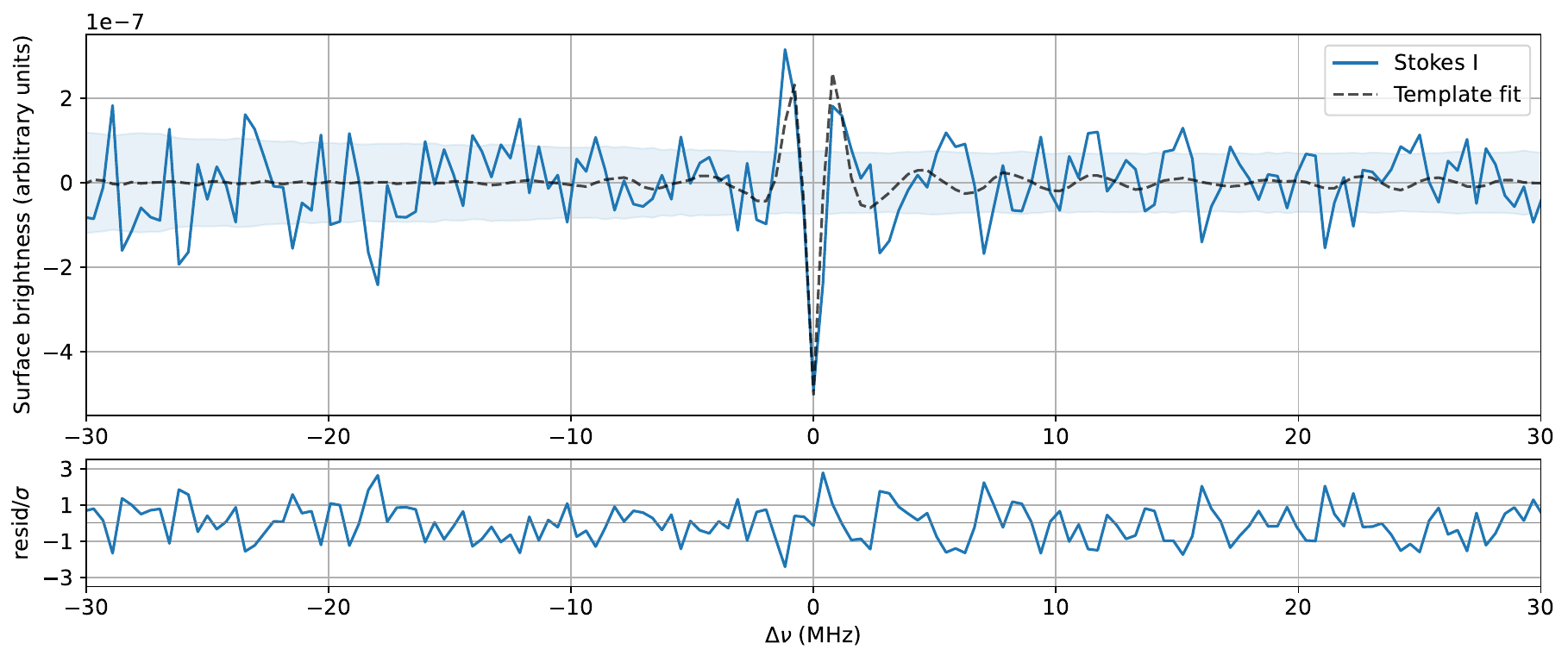}
    \caption{\emph{(Top)} Cross-correlation function of CHIME and eBOSS \lya
    forest data, for the combined X and Y polarisations. An estimate of the
    standard deviation of the background, as described in
    Section~\ref{sec:lya-perm}, is plotted as a shaded region. The black dashed
    line is a template derived from simulations with an amplitude fit to the
    data. \emph{(Bottom)} Residuals normalised by the estimated background
    level.}
    \label{fig:xcorr_fit}
\end{figure*}

\begin{figure}[htb]
    \centfig{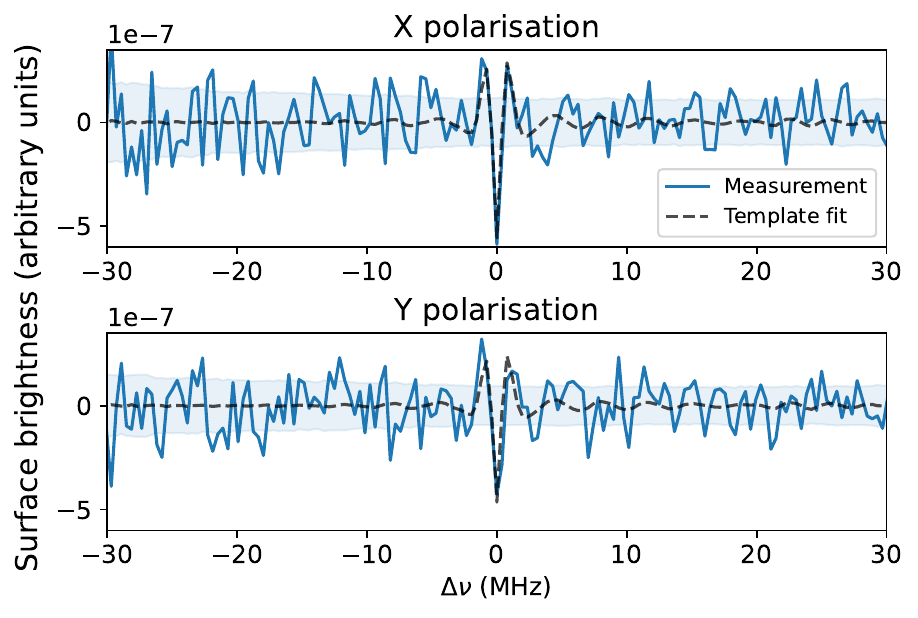}
    \caption{Measured cross-correlation functions for each polarisation, along
    with the corresponding template fit to that polarisation.}
    \label{fig:xcorr_fit_pol}
\end{figure}

The cross-correlation functions of CHIME and the eBOSS \lya forest are presented
in Figure~\ref{fig:xcorr_fit}, where the two polarisations have been combined to
approximate total intensity (they are shown separately in
Figure~\ref{fig:xcorr_fit_pol}). Note that the cross-polar response of the
telescope beam has not been calibrated, so this approximation may include
leakage from polarised emission as well. An estimate of the contamination from
noise and residual foregrounds was derived by cross-correlating the CHIME data
against permutations of the \lya forest spectra, i.e. exchanging the spectrum
measurements between lines of sight so that they are uncorrelated with the CHIME
observations but maintain exactly the same distribution on the sky (this is
described in more detail below). The measurements show a clear excess over the
background around the zero lag bin, with the negative sign characteristic of
this correlation. Also shown in this figure is a signal template derived from
the result of cross-correlating the simulated datasets described in the previous
section. Only the amplitude of the template was fit to the data, as explained
below, and it appears to be broadly consistent with the measurement.

\subsection{\lya Forest Permutations}
\label{sec:lya-perm}

In order to characterise the background fluctuations in the cross-correlation
estimates, we would like to generate a catalogue of \lya forest measurements
with the same properties as the eBOSS one but which is uncorrelated with the
CHIME data. Noise, foreground residuals, and any other systematic effects that
are not correlated with the signal will remain present in the resulting
correlation functions, and we can measure their power. We do this by drawing
random permutations of the \lya forest spectra, while keeping the sample of
line-of-sight directions and quasar redshifts fixed. 1000 such permutations were
generated and correlation functions evaluated for all of them. The standard
deviation over permutations at every spectral separation is shown as the shaded
region in Figure~\ref{fig:xcorr_fit}. Note that correlations between frequency
separations were not considered (i.e. the covariance was assumed to be
diagonal).

Also note that this procedure keeps the selection of lines-of-sight in the CHIME
data fixed, such that if by chance it included data which deviates significantly
from the rest it would be present in every permutation and potentially bias the
estimated variance. To check for this, we generate another set of permutations
where the catalogue positions have also been rotated by $\sim 3$\degree
(approximately one CHIME primary beam width at these wavelengths) in a random
direction. We find that the measured variance in the background fluctuations
from these rotated permutations is indistinguishable from the unrotated set
within the sample noise. Thus, we do not expect that the variance computed from
the unrotated permutations is subject to significant bias due to the fixed set
of lines of sight.

\subsection{Detection Significance}

To assess the significance of the detection, we derive a signal template from
the simulations. Synthetic CHIME and \lya forest data were generated without
adding noise and run through the same filtering and cross-correlation analysis.
The resulting correlation functions, the dashed lines in
Figure~\ref{fig:xcorr_fit_pol}, have a shape that is compact in frequency
separation, a result of the delay filter removing larger scales. These were fit
to the cross-correlation measurements by varying a single amplitude parameter
and minimising the $\chi^2$. Finding the $\Delta\chi^2=1$ bounds in parameter
space provides an estimate for the standard deviation of the amplitude
constraint. The fitted template and residuals are presented in
Figure~\ref{fig:xcorr_fit} for the combined polarisations. We quote the
significance of the detection as the amplitude of the template in units of its
standard deviation from the fit in Table~\ref{tab:signif}.

The fitted amplitude of the template to match the measured signal is a factor of
$\sim4$, i.e. the measurement is about 4 times brighter than the simulation.
Given that the CHIME instrument model used in the simulations is a simple
approximation -- in particular, the model for the beam response which is
essential to accurately normalise the signal amplitude -- we are not confident
in interpreting this as a discrepancy between the physics of the model and the
observations. However, future work that refines both the instrumental
calibration and modelling should allow a useful amplitude to be constrained.

\begin{table}[htb!]
    \centering
    \caption{Detection significance of correlation function template fits. The number of degrees of freedom used to compute $\chi_\nu$ is 254.}
    \begin{tabular*}{\linewidth}{@{\extracolsep{\fill}} c c c c}
        \textbf{measurement} & $\mathbf{\chi^2}$ & $\mathbf{\chi^2_\nu}$ & \textbf{significance} \\
        \midrule
        X pol \hfill & 217.1 & 0.85 & $6.4\sigma$ \\
        Y pol \hfill & 295.8 & 1.16 & $6.5\sigma$ \\
        combined \hfill & 238.5 & 0.94 & $9.1\sigma$ \\
    \end{tabular*}
    \label{tab:signif}
\end{table}
\section{Validation}
\label{sec:valid}

In this section we describe checks that were performed to validate the detection
and background estimate.

\subsection{Impact of Selection Function on Catalogue Permutations}

To estimate the level of background fluctuations in the correlation function
measurements, random permutations of the \lya forest catalogue were drawn and
correlated against the CHIME data, as described in the previous section. By
construction, this method preserves the distribution of background quasars on
the sky and in redshift separately, but if there exist correlations between
these two axes in the selection function, they will be erased by our procedure.
We check for the presence of such a correlation by binning the background
quasars by their angular position onto a HEALPix \citep{healpix} grid of `nside'
32 and computing the average redshift in each bin. The standard deviation across
non-empty bins is $\Delta z < 0.08$, or about 3\% of the average redshift. For
comparison, we shuffled the positions of the quasars, and repeated the binning
calculation, resulting in a typical standard deviation of just under 3\%. There
doesn't appear to be more variation in the redshifts with location on the sky
than would be expected for an uncorrelated distribution and we conclude that
selection effects are unlikely to be important.

\subsection{Even/Odd Days}

For this test, the full set of days of CHIME data that went into the sidereal
stack were divided into two sets according to an even/odd split in chronological
order, and each was stacked in the same way as the full set (see
Section~\ref{sec:data}). The \tcm and foreground signals are the same on every
sidereal day, so they are expected to be common to the even and odd stacks, but
thermal noise and RFI should be uncorrelated between the two. By taking the
difference (and dividing by a factor of two), the common foreground residuals
should cancel and what is left can be compared to our expectation for thermal
noise. Any remaining excess over the noise can be attributed to day-to-day
variations, e.g. RFI or changes to the instrumental response that were not
entirely captured in the daily calibration.

\begin{figure}
    \centfig{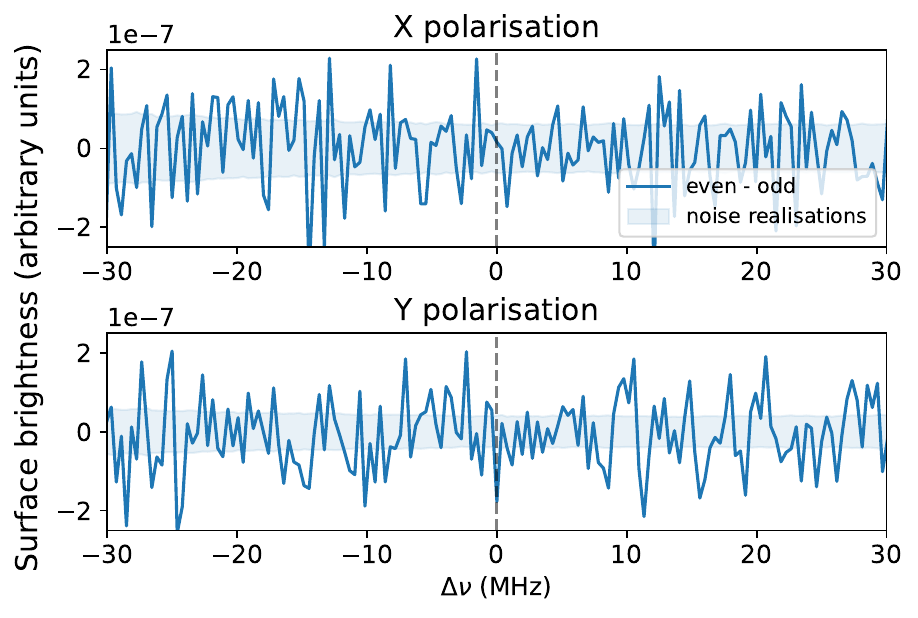}
    \caption{Correlation function evaluated on the difference of even and odd
    days, with the standard deviation derived from noise realisations as a
    shaded region}
    \label{fig:even-odd}
\end{figure}

\begin{figure}
    \centfig{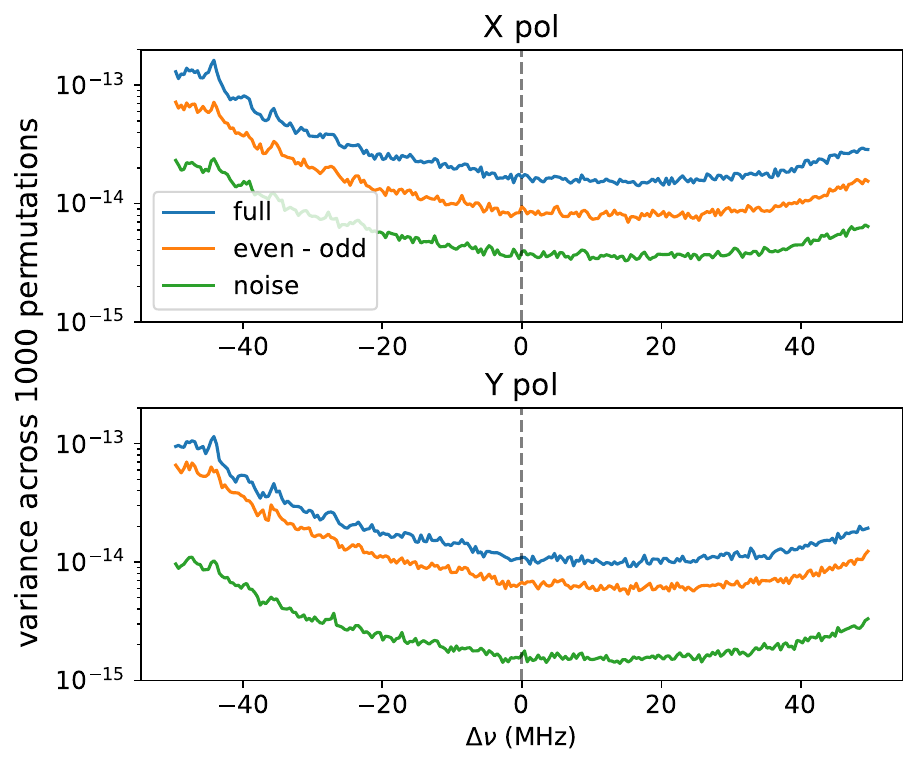}
    \caption{Variance computed across correlation functions estimated on 1000
    permutations of the \lya forest spectra. In blue is the full CHIME data,
    orange is the difference of even and odd days and green is a noise
    realisation. Note that the minimum does not occur at zero separation because
    the maximal overlap of the highest weighted regions in the CHIME and eBOSS
    spectra is not achieved there. Also, the full range of frequency separations
    is shown here, whereas previous plots were cropped to better show the
    signal.}
    \label{fig:noise-compare}
\end{figure}

The correlation functions evaluated on the even/odd difference are shown in
Figure~\ref{fig:even-odd}, along with the expected noise level. Not
surprisingly, there is no visible excess around the zero lag bin, which was
confirmed by fitting the signal template to this data, resulting in amplitude
parameters that are consistent with zero within $1\sigma$. Note that the
background fluctuations appear to have a magnitude in excess of the noise. To
make this comparison more clearly, the correlation function was evaluated for
1000 permutations of the \lya forest spectra for the even-odd difference, as
well as for a CHIME noise realisation. In all cases, the same set of
permutations was used. Figure~\ref{fig:noise-compare} shows the variance across
permutations for these two cases along with the regular data. The ratio between
the variance in the regular analysis and the even-odd difference indicates the
contribution of residual foregrounds to the background fluctuations in the
correlation function. These account for about twice as much variance as the
even-odd difference, which in turn is about 3-5 times as large as the noise.
Note however that the variance is skewed by significant non-Gaussian tails in
the distributions of the pixel values.

\begin{figure*}
    \centfig{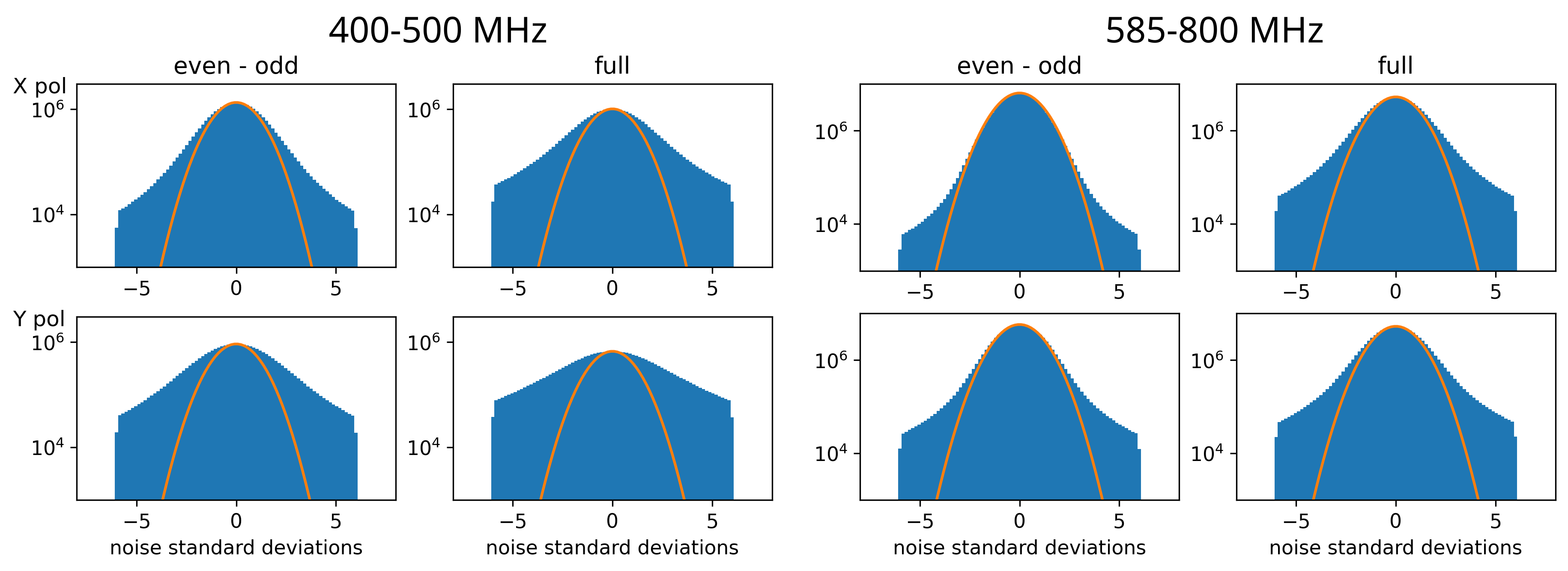}
    \caption{Histogram of pixel values in filtered and masked ringmaps across
    all frequencies used in \emph{(left)} the lower band, 400-500~MHz, and
    \emph{(right)} the upper band, 585-800~MHz. The pixels are normalised to the
    expected noise standard deviation, and a Gaussian curve with a matching
    amplitude is plotted in orange for comparison. Rows show the two
    polarisations, and columns show the difference of even and odd days and the
    average of all days. The effect of the outlier mask can be seen as the sharp
    cutoff at $6\sigma$.}
    \label{fig:even-odd-hist}
\end{figure*}

Figure~\ref{fig:even-odd-hist} shows histograms of the pixels in the filtered
and masked maps across all frequencies that were used in the upper and lower
bands, within a region of the sky restricted roughly to the eBOSS NGC field. In
the even-odd difference of the lower band, the distribution is closer to
Gaussian for values $\lesssim 2\sigma$, but large tails remain. The histograms
of the upper band are much closer to Gaussian for $\lesssim 3\sigma$, especially
in the even-odd difference.

The analysis of the difference of even and odd days in the lower band suggests
that
\begin{enumerate}
    \item There is a large excess of variance in the maps above the thermal
    noise (by a factor $\sim 6-10$).
    \item This excess is not fully accounted for by features that are repeated
    day-to-day like celestial foregrounds (the variance is only $\sim$halved in
    the even-odd difference).
    \item The excess above the noise is non-Gaussian, with large tails in the
    distribution of map pixels.
    \item The distribution is significantly less Gaussian in the lower band than
    in the upper band, especially in the even-odd difference.
\end{enumerate}
A possible explanation for this is that the lower band is more contaminated than
the upper band by unflagged RFI -- which does not repeat day-to-day and
contributes in a non-Gaussian-distributed way. Another contributing effect could
also be errors in the daily calibration, which would have a similar signature.
Both of these factors have been observed in other internal CHIME analyses of the
lower band. Identifying the source of this contamination and removing it will
likely require further investigation of the individual days, before they are
averaged into a sidereal stack. While these residuals are concerning and finding
their origin in order to remove them will enable more sensitive analyses, their
presence has not impeded the detection in cross-correlation with the \lya forest
reported in this work.
\section{Conclusion}
\label{sec:conc}

This work reports on the first detection of \tcm emission at redshifts $z >
1.5$, obtained by cross-correlating CHIME data from 88 days in the $400-500$~MHz
band with the DR16 eBOSS measurements of the \lya forest. To do so, we extended
the analysis methods described in \citet{chime_collaboration_detection_2022} to
the lower quarter of the CHIME frequency band, and developed a cross-correlation
method to combine this data with the \lya forest as a tracer of large-scale
structure.

This demonstrates that CHIME is able to detect emission from cosmological HI in
the entire redshift range available to it, but it also highlights the difficulty
of separating this signal from the extremely bright foregrounds. Even in
cross-correlation and after aggressive delay filtering, excess variance from
foregrounds and other sources is found to be about $6-10$ times above thermal
noise. This level of contamination nevertheless allows for a $\sim 9\sigma$
detection. The shape of the measured correlation function is dominated by the
effect of the foreground filter, and although we have not attempted to interpret
it, we have found that a simple physical model agrees with the measurement well
enough to use this model to quantify the detection significance. The amplitude
of the cross-correlation should in principle tell us something about the physics
of the HI that sources the \lya forest and its relationship to the
high-HI-density systems primarily probed with \tcm intensity mapping, but
interpreting its measurement will require more sophisticated modelling of the
\lya and \tcm signals. This will be the subject of future work.

Improvements to the CHIME calibration and analysis pipelines, as well as the
inclusion of several more years of data that have yet to be processed, will lead
to future improvements in the constraining power of the analysis. Significant
residuals in day-to-day jacknives also suggest that the lower band considered in
this work is more contaminated than the upper band that was reported on
previously, and may especially benefit from improved RFI flagging. In the short
term, this might reduce the number of frequencies that need to be flagged and
allow for a less restrictive delay cut, enabling access to information from
scales larger than the $k_\parallel \gtrsim 0.13\ \text{Mpc}^{-1}$ that remain
in this analysis. These improvements will position CHIME's dataset as a powerful
source of cross- and auto-correlation measurements.
\section*{Acknowledgements}

We thank Meiling Deng for her substantial contributions to the
instrument design and analysis of CHIME.

We thank the Dominion Radio Astrophysical Observatory, operated by the National
Research Council Canada, for gracious hospitality and expertise. The DRAO is
situated on the traditional, ancestral, and unceded territory of the Syilx
Okanagan people. We are fortunate to live and work on these lands.

CHIME is funded by grants from the Canada Foundation for Innovation (CFI) 2012
Leading Edge Fund (Project 31170), the CFI 2015 Innovation Fund (Project 33213),
and by contributions from the provinces of British Columbia, Québec, and
Ontario. Long-term data storage and computational support for analysis is
provided by WestGrid, SciNet and the Digital Research Alliance of Canada, and we
thank their staff for flexibility and technical expertise that has been
essential to this work.

Additional support was provided by the University of British Columbia, McGill
University, and the University of Toronto. CHIME also benefits from NSERC
Discovery Grants to several researchers, funding from the Canadian Institute for
Advanced Research (CIFAR), from Canada Research Chairs, from the FRQNT Centre de
Recherche en Astrophysique du Québec (CRAQ) and from the Dunlap Institute for
Astronomy and Astrophysics at the University of Toronto, which is funded through
an endowment established by the David Dunlap family. This material is partly
based on work supported by the NSF through grants (2008031) (2006911) and
(2006548) and by the Perimeter Institute for Theoretical Physics, which in turn
is supported by the Government of Canada through Industry Canada and by the
Province of Ontario through the Ministry of Research and Innovation.

We thank the Sloan Digital Sky Survey and eBOSS collaborations for publicly
releasing the \lya forest catalogs used in this work. Funding for the Sloan
Digital Sky Survey IV has been provided by the Alfred P. Sloan Foundation, the
U.S. Department of Energy Office of Science, and the Participating Institutions.
SDSSIV acknowledges support and resources from the Center for High Performance
Computing at the University of Utah. The SDSS website is \url{www.sdss.org}.

\software{
bitshuffle \citep{2015bitshuffle},
CAMB \citep{lewis1999},
caput \citep{caput},
ch\_pipeline \citep{ch_pipeline},
cora \citep{cora},
Cython \citep{Cython},
draco \citep{draco},
driftscan \citep{driftscan},
hankl \citep{karamanis2021},
h5py \citep{h5py},
HDF5 \citep{HDF5},
HEALPix \citep{healpix},
healpy \citep{healpy},
Matplotlib \citep{Matplotlib},
mpi4py \citep{mpi4py},
NumPy \citep{NumPy},
OpenMPI \citep{OpenMPI},
pandas \citep{pandas,pandas_paper},
peewee \citep{Peewee},
SciPy \citep{SciPy},
Skyfield \citep{Skyfield},
}


\bibliography{paper}{}
\bibliographystyle{aasjournal}

\end{document}